\documentclass[letterpaper]{article} 
\usepackage{aaai2026}  
\usepackage{times}  
\usepackage{helvet}  
\usepackage{courier}  
\usepackage[hyphens]{url}  
\usepackage{graphicx} 
\usepackage{natbib}  
\usepackage{caption} 
\usepackage{algorithm}
\usepackage{algorithmic}
\usepackage{amsmath}   
\usepackage{amssymb}   
\usepackage{multirow}  
\usepackage{newfloat}
\usepackage{listings}
\DeclareCaptionStyle{ruled}{labelfont=normalfont,labelsep=colon,strut=off} 
\floatstyle{ruled}
\newfloat{listing}{tb}{lst}{}
\floatname{listing}{Listing}
\title{HiveMind: Contribution-Guided Online Prompt Optimization of LLM Multi-Agent Systems}
\author{
    Yihan Xia\textsuperscript{\rm 1},
    Taotao Wang\textsuperscript{\rm 1,†},
    Shengli Zhang\textsuperscript{\rm 1},
    Zhangyuhua Weng\textsuperscript{\rm 1},
    Bin Cao\textsuperscript{\rm 2},
    Soung Chang Liew\textsuperscript{\rm 3}
    \thanks{\textsuperscript{†} Corresponding author.}
}
\affiliations{
    \textsuperscript{\rm 1}College of Electronic and Information Engineering, Shenzhen University\\
    \textsuperscript{\rm 2}School of Information and Communication Engineering, Beijing University of Posts and Telecommunications\\
    \textsuperscript{\rm 3}Department of Information Engineering, The Chinese University of Hong Kong\\
    xiayihan2023@email.szu.edu.cn, ttwang@szu.edu.cn, zsl@szu.edu.cn, \\2410043007@mails.szu.edu.cn, caobin@bupt.edu.cn, soung@ie.cuhk.edu.hk
 
}

\begin{document}

\maketitle

\begin{abstract}
Recent advances in LLM-based multi-agent systems have demonstrated remarkable capabilities in complex decision-making scenarios such as financial trading and software engineering. However, evaluating each individual agent's effectiveness and online optimization of underperforming agents remain open challenges. To address these issues, we present HiveMind, a self-adaptive framework designed to optimize LLM multi-agent collaboration through contribution analysis. At its core, HiveMind introduces Contribution-Guided Online Prompt Optimization (CG-OPO), which autonomously refines agent prompts based on their quantified contributions. We first propose the Shapley value as a grounded metric to quantify each agent's contribution, thereby identifying underperforming agents in a principled manner for automated prompt refinement. To overcome the computational complexity of the classical Shapley value, we present DAG-Shapley, a novel and efficient attribution algorithm that leverages the inherent Directed Acyclic Graph structure of the agent workflow to axiomatically prune non-viable coalitions. By hierarchically reusing intermediate outputs of agents in the DAG, our method further reduces redundant computations, and achieving substantial cost savings without compromising the theoretical guarantees of Shapley values. Evaluated in a multi-agent stock-trading scenario, HiveMind achieves superior performance compared to static baselines. Notably, DAG-Shapley reduces LLM calls by over 80\% while maintaining attribution accuracy comparable to full Shapley values, establishing a new standard for efficient credit assignment and enabling scalable, real-world optimization of multi-agent collaboration.
\end{abstract}


\section{Introduction}

Large Language Models (LLMs) have revolutionized Multi-Agent Systems (MAS), enabling the creation of sophisticated agent teams capable of addressing complex collaborative tasks across diverse domains, such as financial trading, automated software development, and scientific discovery. Despite their remarkable potential, deploying these systems in dynamic, real-world environments introduces critical challenges, particularly in maintaining optimal performance through continuous, autonomous adaptation.

Two fundamental issues arise in this context. First, optimizing underperforming agents autonomously during runtime is essential for adapting to dynamic environments but remains an open problem. Second, accurately attributing individual agents' contributions to the overall system performance is a significant challenge, especially in complex workflows where interactions between agents are non-trivial. Current solutions often rely on static agent configurations, manually crafted prompts, and periodic updates. While this approach may suffice in stable environments, it struggles in dynamic scenarios where agent roles and requirements evolve rapidly, necessitating real-time adaptation.

To address these challenges, we propose \textbf{HiveMind}, a self-adaptive framework designed to optimize LLM-based multi-agent collaboration through autonomous refinement and contribution analysis. At its core, HiveMind introduces \textbf{Contribution-Guided Online Prompt Optimization (CG-OPO)}, a closed-loop optimization mechanism that continuously refines agent prompts based on their quantified contributions. By autonomously monitoring agent performance, CG-OPO identifies underperforming agents and initiates targeted prompt refinements through an LLM-driven reflection process, enabling dynamic system improvement during runtime. A key enabler of CG-OPO is \textbf{DAG-Shapley}, a novel and efficient algorithm for contribution measurement. DAG-Shapley leverages the Directed Acyclic Graph (DAG) structure inherent in agent workflows to compute Shapley values, a principled metric for quantifying individual contributions. By axiomatically pruning non-viable coalitions and hierarchically reusing intermediate outputs, DAG-Shapley significantly reduces computational complexity compared to classical Shapley value calculations, making real-time attribution feasible without compromising theoretical guarantees.

HiveMind operates as a self-adaptive cycle of diagnosis and optimization: DAG-Shapley provides rapid and accurate identification of performance bottlenecks, while CG-OPO executes targeted adjustments to enhance agent collaboration. This framework eliminates the need for human intervention, ensuring scalable and robust MAS performance in dynamic environments. Our primary contributions are as follows:
\begin{itemize}
\item \textbf{Dynamic Autonomous Optimization Framework}: We introduce CG-OPO, a mechanism that enables continuous refinement of agent prompts based on quantified contributions, ensuring sustained system improvement during runtime.
\item \textbf{Efficient Contribution Measurement Algorithm}: We propose DAG-Shapley, a computationally efficient algorithm for real-time attribution in DAG-structured agent systems, reducing LLM calls by over 80\% compared to classical Shapley value calculations.
\item \textbf{Integrated Self-Adaptive Framework}: HiveMind combines CG-OPO and DAG-Shapley into a cohesive framework for building scalable, self-optimizing agent teams.
\end{itemize}

Evaluated in a multi-agent stock-trading scenario, HiveMind demonstrates superior performance compared to static baselines, achieving efficient credit assignment and scalable optimization of multi-agent collaboration. By addressing the challenges of dynamic optimization and real-time attribution, HiveMind establishes a new standard for MAS, paving the way for robust applications in complex, evolving environments.

\section{Related Work}

Our work builds upon and contributes to three primary areas of research: LLM-based multi-agent systems, credit assignment in cooperative games, and automated prompt engineering.

\subsection{LLM-Based Multi-Agent Systems}

The concept of multi-agent systems has a rich history, but the advent of LLMs has catalyzed a new wave of research into more capable and flexible agent architectures \cite{wang2024survey, zhang2023building}. Recent works have demonstrated the potential of LLM-based agents for complex, collaborative tasks. In software development, frameworks like MetaGPT \cite{hong2023metagpt} and ChatDev \cite{qian2023chatdev} simulate entire software companies with specialized agent roles. For social science and human behavior studies, systems like Generative Agents \cite{park2023generative} create believable simulacra of human interaction in sandbox environments. Beyond specific domains, general-purpose collaborative frameworks such as AgentVerse \cite{chen2023agentverse} and AutoGen \cite{wu2023autogen} have been proposed to facilitate cooperation among LLM agents to solve a wide array of complex problems. Recent advances in multi-modal information processing have also explored efficient hashing techniques for handling heterogeneous data types in collaborative systems \cite{11164571}. While these systems showcase impressive capabilities, the coordination and optimization of agent collaboration often rely on predefined heuristics or manual intervention. Our work moves beyond this by proposing a framework for autonomous, online optimization of the agent workflow itself.

\subsection{Credit Assignment}

The problem of fairly attributing a collective outcome to individual contributors is a central theme in both reinforcement learning \cite{sutton2018reinforcement} and cooperative game theory. The Shapley value \cite{shapley1953value} remains the gold standard for fair credit assignment due to its strong axiomatic guarantees. However, its exponential complexity has spurred a long line of research into approximation methods. In multi-agent reinforcement learning, credit assignment is often addressed through value function factorization, as seen in influential works like VDN \cite{sunehag2017value} and QMIX \cite{rashid2018qmix}. Our work, DAG-Shapley, falls into the category of structural credit assignment, we are the first to leverage the inherent Directed Acyclic Graph (DAG) structure of LLM-based agent workflows to create a computationally efficient and accurate credit assignment mechanism specifically for LLM-based systems.

\subsection{Automated Prompt Engineering}

The performance of LLMs is highly sensitive to the quality of their prompts. This has led to a surge in research on automated prompt engineering, or ``prompt optimization.'' Foundational prompting strategies that elicit complex reasoning, such as Chain-of-Thought (CoT) \cite{wei2022chain} and Tree-of-Thoughts (ToT) \cite{yao2023tree}, have demonstrated the profound impact of prompt structure. Building on this, automated methods have emerged to discover optimal prompts. These include discrete optimization methods like Automatic Prompt Engineer (APE) \cite{zhou2022large}, gradient-free optimization \cite{pryzant2023automatic}, and methods using reinforcement learning to refine prompts \cite{deng2022rlprompt}. Complementary research in vision-language pre-training has demonstrated the importance of semantic completion learning for enhancing multi-modal understanding capabilities \cite{tu2025global}. Our Contribution-Guided Online Prompt Optimization (CG-OPO) framework extends this concept to the multi-agent setting. It represents a novel approach where the optimization of individual agent prompts is guided by a global, system-level assessment of each agent's contribution, creating a direct, performance-driven link between collaborative success and prompt refinement.

\begin{figure*}
\centering
\includegraphics[width=0.8\textwidth]{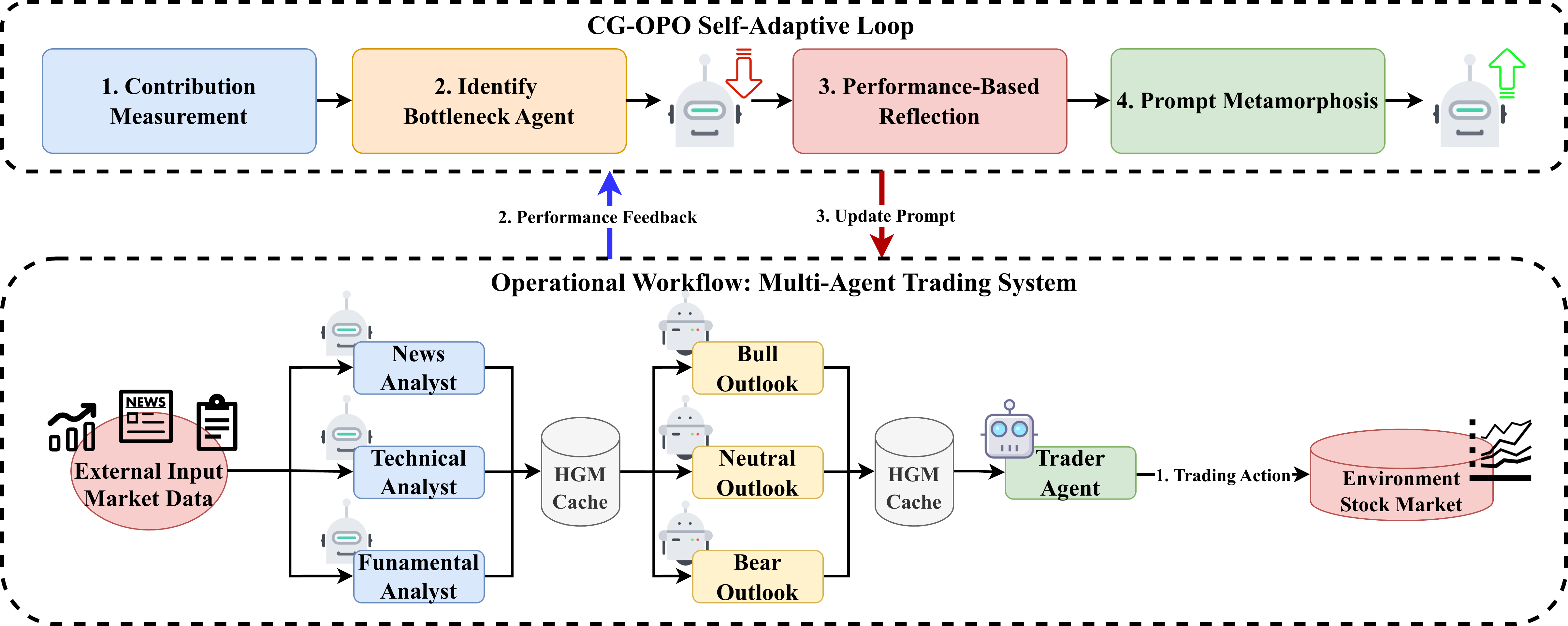}
\caption{Overall system architecture showing the closed-loop optimization cycle.}
\label{fig:system_architecture}
\end{figure*}

\section{System Formalization}

This section details HiveMind, our self-adaptive optimization framework for LLM-based multi-agent systems. The framework addresses two fundamental challenges: autonomous optimization of underperforming agents during runtime and accurate attribution of individual agent contributions to overall system performance. HiveMind introduces two key innovations: Contribution-Guided Online Prompt Optimization (CG-OPO), a closed-loop optimization mechanism that continuously refines agent prompts based on quantified contributions; and DAG-Shapley, an efficient algorithm for contribution measurement that leverages Directed Acyclic Graph structure to compute Shapley values with dramatically reduced computational complexity.

HiveMind operates as a closed-loop, periodic optimization system that continuously improves collective performance through autonomous adaptation. The framework consists of an iterative cycle: performance evaluation $\rightarrow$ contribution measurement $\rightarrow$ bottleneck identification $\rightarrow$ prompt optimization $\rightarrow$ system update. At its core, CG-OPO provides targeted agent enhancement guided by precise contribution measurement from Shapley value. This combination eliminates manual intervention requirements while ensuring scalable optimization in dynamic environments. Figure~\ref{fig:system_architecture} illustrates the primary components and their data flow within this continuous optimization loop, where each iteration refines individual agent behavior based on measured contributions to overall system performance.

\subsection{System Architecture and Information Flow}

To systematically manage agent interactions and enable principled contribution analysis, we model our Multi-Agent System (MAS) as a Directed Acyclic Graph (DAG). This architectural choice is motivated by three key observations: (1) many collaborative tasks exhibit natural information flow hierarchies that preclude cycles, (2) acyclic structures enable efficient topological scheduling and prevent information loops that can destabilize learning, and (3) DAG constraints facilitate rigorous credit assignment through game-theoretic frameworks \cite{shapley1953value,li2024shapley}. Our approach builds upon recent advances in DAG-based multi-agent coordination \cite{qian2024scaling} while addressing the specific challenges of financial decision-making under uncertainty.

\subsubsection{Graph Formalization and Constraints}

Formally, we define the system as a graph $\mathcal{G} = (\mathcal{V}, \mathcal{E})$, where $\mathcal{V} = \{a_1, a_2, \ldots, a_N\}$ is the set of $N$ agents, and $\mathcal{E} \subseteq \mathcal{V} \times \mathcal{V}$ is the set of directed edges. An edge $(a_i, a_j) \in \mathcal{E}$ signifies that the output of agent $a_i$ is a required input for agent $a_j$. The acyclicity constraint ensures that there exists a topological ordering $\pi: \mathcal{V} \rightarrow \{1, 2, \ldots, N\}$ such that for all $(a_i, a_j) \in \mathcal{E}$, we have $\pi(a_i) < \pi(a_j)$.

Within this graph, we distinguish critical subsets of nodes:
\begin{itemize}
\item \textbf{Source nodes}: $\mathcal{V}_{\text{source}} = \{v \in \mathcal{V} : \text{in-degree}(v) = 0\}$, which process external information
\item \textbf{Sink node}: $a_T \in \mathcal{V}$ such that $\text{out-degree}(a_T) = 0$, which produces the system's final output
\item \textbf{Intermediate nodes}: $\mathcal{V}_{\text{intermediate}} = \mathcal{V} \setminus (\mathcal{V}_{\text{source}} \cup \{a_T\})$
\end{itemize}

\subsubsection{Access Control and Data Dependencies}

The DAG structure enforces strict information flow constraints through a formal access control mechanism \cite{myerson1977graphs}. Let $\mathcal{D}$ represent the universe of available data, and define an information access function $I: \mathcal{V} \rightarrow 2^{\mathcal{D}}$ that maps each agent to its accessible data subset. For any agent $a_i$, its information set $I(a_i)$ includes:
\begin{equation}
I(a_i) = \mathcal{D}_{\text{external}}^{(i)} \cup \bigcup_{(a_j, a_i) \in \mathcal{E}} O(a_j)
\end{equation}
where $\mathcal{D}_{\text{external}}^{(i)}$ represents external data directly accessible to agent $a_i$, and $O(a_j)$ denotes the output of agent $a_j$. This formalization ensures that information flows strictly according to the DAG topology, preventing unauthorized data access that could compromise the system's hierarchical structure.

\subsubsection{Structural Benefits and Complexity}

The DAG constraint provides several computational and organizational advantages:

\begin{enumerate}
\item \textbf{Efficient Scheduling}: The topological ordering enables parallel execution within layers while maintaining dependencies, achieving optimal scheduling complexity $O(|\mathcal{V}| + |\mathcal{E}|)$.

\item \textbf{Bounded Computation}: The acyclic property guarantees termination and prevents infinite loops in iterative processes, ensuring system stability.

\item \textbf{Hierarchical Learning}: The structured information flow enables layer-wise optimization and facilitates credit assignment through backward propagation of gradients or game-theoretic value distribution.
\end{enumerate}

\subsubsection{Implementation and Agent Roles}

To ground this abstraction, we instantiate our framework within a multi-agent financial trading system designed for stock market analysis and decision-making, a domain where multi-agent approaches have demonstrated significant promise \cite{xiao2024tradingagents,tran2025multi}. The system comprises seven agents organized into three distinct hierarchical layers:

\textbf{Analysis Layer} ($\mathcal{V}_{\text{source}}$): This layer consists of three independent specialist agents:
\begin{itemize}
\item \textbf{News Analyst Agent (NAA)}: Processes news feeds and sentiment data
\item \textbf{Technical Analyst Agent (TAA)}: Analyzes price history and technical indicators  
\item \textbf{Fundamental Analyst Agent (FAA)}: Evaluates financial statements and company metrics
\end{itemize}

Each agent processes raw, heterogeneous market data in parallel to produce focused analytical summaries, forming the source nodes of our DAG.

\textbf{Outlook Layer} ($\mathcal{V}_{\text{intermediate}}$): This layer contains three agents that synthesize outputs from the Analysis Layer:
\begin{itemize}
\item \textbf{Bullish Outlook Agent (BOA)}: Aggregates positive market signals
\item \textbf{Bearish Outlook Agent (BeOA)}: Consolidates negative market indicators
\item \textbf{Neutral Outlook Agent (NOA)}: Balances conflicting signals
\end{itemize}

These agents are explicitly prohibited from accessing raw market data through the information flow constraints defined by $I(\cdot)$, forcing reliance solely on structured analyses from the previous layer.

\textbf{Decision Layer} (Sink Node): The \textbf{Trader Agent (TRA)} constitutes the final layer, consuming competing outlooks to produce actionable trading decisions. This agent serves as the system's sink node $a_T$.

The integration of Shapley value theory into this DAG-based framework provides a principled approach to contribution assessment that leverages the structured information flow \cite{myerson1977graphs}. This combination enables both efficient execution and fair credit assignment, making it particularly suitable for collaborative decision-making tasks where individual agent contributions must be quantified and optimized.

\subsection{Contribution-Guided Online Prompt Optimization}

In multi-agent systems, evaluating each individual agent's effectiveness and performing online optimization of underperforming agents remain open challenges.

To address these challenges, we introduce \textbf{Contribution-Guided Online Prompt Optimization (CG-OPO)}, a systematic framework that autonomously transforms contribution measurement scores into targeted agent improvements. CG-OPO establishes a principled feedback mechanism utilizing game-theoretic attribution to concentrate optimization on underperforming components while preserving well-functioning agents. By integrating Shapley value-based attribution with automated prompt optimization techniques \cite{li2025survey,pryzant2023automatic}, CG-OPO addresses fundamental system bottlenecks, enabling scalable autonomous improvement in DAG-structured multi-agent architectures

In our framework, the optimization is iterative. We use $t$ to denote the current optimization cycle or time step (e.g., a weekly evaluation period), during which historical data $\mathcal{H}_t$ is collected.

\subsubsection{Four-Stage Optimization Pipeline}

CG-OPO operates through four sequential stages:
\begin{itemize}
    \item \textbf{Contribution Measurement}: Evaluates individual agent performance using game-theoretic attribution
    \item \textbf{Identify Bottleneck Agent}: Selects the lowest-performing agent for optimization
    \item \textbf{Performance-Based Reflection}: Analyzes failure and success cases via meta-optimizer to extract improvement insights
    \item \textbf{Prompt Metamorphosis}: Integrates learned lessons into the target agent's prompt for enhanced performance
\end{itemize}

The complete optimization process is formalized in Algorithm~\ref{alg:cg_opo}.

\subsubsection{Stage 1: Contribution Measurement}

The optimization cycle begins with comprehensive evaluation of each agent's contribution to system performance. For each agent $a_i \in \mathcal{V}$, we compute the Shapley value $\phi_i(t)$ representing agent $i$'s contribution to the system's overall performance, based on the Sharpe ratio of coalition subsets \cite{hua2025shapley,liang2024asynchronous}. The Shapley value for agent $i$ is defined as:
\begin{equation}
\phi_i(v) = \sum_{S \subseteq \mathcal{V} \setminus \{a_i\}} \frac{|S|!(N - |S| - 1)!}{N!} [v(S \cup \{a_i\}) - v(S)]
\end{equation}
where $v(S)$ represents the coalition performance measured by the Sharpe ratio:
\begin{equation}
v(S) = \text{Sharpe}(S) = \frac{\mathbb{E}[R_S] - R_f}{\sigma(R_S)}
\end{equation}
where $R_S$ is the return generated by coalition $S$, $R_f$ is the risk-free rate, and $\sigma(R_S)$ is the standard deviation of coalition returns.

\subsubsection{Stage 2: Identify Bottleneck Agent}

Based on the contribution scores computed in Stage 1, the system identifies the target agent for optimization:
\begin{equation}
a^* = \arg\min_{a_i \in \mathcal{V}} \phi_i(t)
\end{equation}
Optimization is triggered only if $\phi_{a^*}(t) < \tau_{\text{threshold}}$, focusing computational resources on genuine bottlenecks while preserving well-performing agents, as shown in Algorithm~\ref{alg:cg_opo}.

\subsubsection{Stage 3: Performance-Based Reflection}

When a bottleneck is identified, the system extracts failure cases $\mathcal{F}^* = \{(s, a, r) \in \mathcal{H}_t(a^*) : r < 0\}$ and success cases $\mathcal{S}^* = \{(s, a, r) \in \mathcal{H}_t(a^*) : r \geq 0\}$ from the target agent's historical performance data $\mathcal{H}_t(a^*)$. A meta-optimizer analyzes these cases to generate actionable lessons $\mathcal{L}_t$ that identify systematic weaknesses and successful patterns through conceptual verbal reinforcement \cite{yu2024fincon}. This approach leverages collaborative agent interaction and iterative refinement of analysis to enhance reasoning capabilities, particularly effective in financial decision-making contexts. The reflective learning mechanism draws upon meta-optimization principles to systematically improve agent performance. The detailed reflection process is presented in Algorithm~\ref{alg:reflection}.

\subsubsection{Stage 4: Prompt Metamorphosis}

The final stage integrates the extracted lessons into the target agent's prompt through structured augmentation, following principles from automated prompt engineering and prompt evolution techniques:
\begin{equation}
P_{t+1}(a^*) = P_t(a^*) \oplus \mathcal{L}_t
\end{equation}
where $\oplus$ denotes concatenation that preserves the agent's core functionality while incorporating experience-based guidance. This approach ensures incremental improvement without disrupting established capabilities, as implemented in Algorithm~\ref{alg:cg_opo}.

\begin{algorithm}[tb]
\caption{Contribution-Guided Online Prompt Optimization (CG-OPO)}
\label{alg:cg_opo}
\textbf{Input}: Agent set $\mathcal{V}$, contribution scores $\{\phi_i(t)\}_{i=1}^n$, weekly I/O data $\mathcal{H}_t$\\
\textbf{Output}: Updated agent prompts for cycle $t+1$
\begin{algorithmic}[1]
\STATE $a^* \leftarrow \arg\min_{a_i \in \mathcal{V}} \phi_i(t)$ // Bottleneck identification
\STATE $\mathcal{F}^* \leftarrow f_{\text{extract}}(\mathcal{H}_t, a^*)$ // Collect performance feedback
\STATE $\Pi_{\text{reflect}} \leftarrow f_{\text{prompt}}(a^*, \phi_{a^*}(t), \mathcal{F}^*)$  // Performance-Based Reflection
\STATE $\mathcal{L}_t \leftarrow f_{\text{reflect}}(\Pi_{\text{reflect}})$ // Generate lessons learned
\STATE $P_{t+1}(a^*) \leftarrow P_t(a^*) \oplus f_{\text{format}}(\mathcal{L}_t)$ // Prompt metamorphosis

\STATE \textbf{return} $\{P_{t+1}(a_i)\}_{i=1}^n$
\end{algorithmic}
\end{algorithm}

\begin{algorithm}[tb]
\caption{Performance-Based Reflection Process}
\label{alg:reflection}
\textbf{Input}: Target agent $a^*$, historical data $\mathcal{H}_t(a^*)$, contribution score $\phi_{a^*}(t)$\\
\textbf{Output}: Lessons learned $\mathcal{L}_t$
\begin{algorithmic}[1]
\STATE $\mathcal{F} \leftarrow \{(s, a, r) \in \mathcal{H}_t(a^*) : r < 0\}$ // Extract failure cases
\STATE $\mathcal{S} \leftarrow \{(s, a, r) \in \mathcal{H}_t(a^*) : r \geq 0\}$ // Extract success cases
\STATE $\Pi_{\text{context}} \leftarrow f_{\text{format}}(\mathcal{F}, \mathcal{S}, \phi_{a^*}(t))$
\STATE $\Pi_{\text{reflect}} \leftarrow \Pi_{\text{template}} \oplus \Pi_{\text{context}}$
\STATE $\mathcal{L}_t \leftarrow f_{\text{optimize}}(\Pi_{\text{reflect}})$
\STATE \textbf{return} $\mathcal{L}_t$
\end{algorithmic}
\end{algorithm}

This creates a complete self-improvement mechanism where contribution measurement enables accurate targeting of genuine bottlenecks, while CG-OPO provides responsive optimization that adapts to changing system dynamics through continuous monitoring and validation of improvements.

\subsection{DAG-Shapley: Efficient Contribution Measurement}

CG-OPO requires precise computation of contribution scores $\phi_i(t)$ for real-time bottleneck identification. While exact Shapley value computation provides optimal attribution accuracy \cite{shapley1953value}, the exponential complexity of evaluating $2^n$ coalitions necessitates efficient algorithms for operational deployment in LLM-based systems where each coalition evaluation requires expensive API calls \cite{hua2025shapley}.

We introduce DAG-Shapley, exploiting DAG structure to achieve exact Shapley computation with enhanced efficiency through coalition space reduction and hierarchical memoization \cite{qian2024scaling,myerson1977graphs,li2020myerson}. The directed nature of our agent workflow aligns with directed graph games theory, where asymmetric information flow and hierarchical relationships naturally emerge. DAG-Shapley operates through two optimization stages: (1) Coalition Space Pruning eliminates structurally invalid coalitions, and (2) Generalized Hierarchical Memoization (GHM) caches functionally equivalent computations.

\subsubsection{Coalition Space Pruning}

Coalition space pruning leverages the DAG structure to eliminate coalitions incapable of producing system outputs. The directed graph framework \cite{li2020myerson} provides theoretical foundation for analyzing coalition formation in hierarchical structures with asymmetric relationships. 

Given workflow structure $\mathcal{G} = (\mathcal{V}, \mathcal{E})$ with source agents $\mathcal{V}_{\text{source}}$ and sink agent $a_T$, coalition viability requires three conditions:
\begin{align}
\text{HasTrader}(S) &\triangleq a_T \in S \label{eq:has_trader}\\
\text{HasSource}(S) &\triangleq S \cap \mathcal{V}_{\text{source}} \neq \emptyset \label{eq:has_source}\\
\text{Connected}(S) &\triangleq \exists v \in (\mathcal{V}_{\text{source}} \cap S) : v \rightsquigarrow a_T \text{ in } \mathcal{G}|_S \label{eq:connected}
\end{align}
A coalition $S$ is considered viable if it meets these conditions. We define a viability function:
\begin{equation}
\text{Viable}(S) \triangleq \text{HasTrader}(S) \land \text{HasSource}(S) \land \text{Connected}(S)
\end{equation}
For non-viable coalitions where $\text{Viable}(S)$ is false, their value is zero ($v(S) = 0$), enabling systematic elimination while preserving Shapley exactness \cite{myerson1977graphs}. The pruning process is formalized in Algorithm~\ref{alg:coalition_pruning}, reducing evaluations from 128 to 49 coalitions for the 7-agent system.

\begin{algorithm}[tb]
\caption{Coalition Space Pruning}
\label{alg:coalition_pruning}
\textbf{Input}: Agent set $\mathcal{V}$, source agents $\mathcal{V}_{\text{source}}$, trader $a_T$, DAG $\mathcal{G}$\\
\textbf{Output}: Viable coalitions $\mathcal{C}_{\text{viable}}$
\begin{algorithmic}[1]
\STATE $\mathcal{C}_{\text{viable}} \leftarrow \emptyset$
\FOR{each coalition $S \in 2^{\mathcal{V}}$}
    \IF{$a_T \notin S$} 
        \STATE \textbf{continue} // Endpoint Rule
    \ENDIF 
    \IF{$S \cap \mathcal{V}_{\text{source}} = \emptyset$} 
        \STATE \textbf{continue} // Source Rule
    \ENDIF 
    \IF{$\nexists v \in \mathcal{V}_{\text{source}} \cap S : v \rightsquigarrow a_T$ in $\mathcal{G}|_S$} 
        \STATE \textbf{continue} // Connectivity Rule
    \ENDIF 
    \STATE $\mathcal{C}_{\text{viable}} \leftarrow \mathcal{C}_{\text{viable}} \cup \{S\}$
\ENDFOR
\STATE \textbf{return} $\mathcal{C}_{\text{viable}}$
\end{algorithmic}
\end{algorithm}

\subsubsection{Generalized Hierarchical Memoization}

The layered architecture enables computational optimization through memoization of functionally equivalent computations. When different coalitions produce identical layer input configurations, downstream computations are functionally equivalent, enabling efficient caching and reuse strategies.

For coalition $S^{(k)} \in \mathcal{C}_{\text{viable}}$, the layer input configuration for layer $\mathcal{L}_i$ is:
\begin{equation}
\mathcal{I}_i^{(k)} = \bigcup_{j=1}^{i-1} (S^{(k)} \cap \mathcal{L}_j)
\end{equation}
Under functional determinism:
\begin{equation}
\mathcal{I}_i^{(k)} = \mathcal{I}_i^{(k')} \Rightarrow O_a(\mathcal{I}_i^{(k)}) = O_a(\mathcal{I}_i^{(k')})
\end{equation}
GHM computes each unique configuration once and reuses results across coalitions, as detailed in Algorithm~\ref{alg:ghm}. The computational cost becomes:
\begin{equation}
\text{Cost}_{\text{GHM}} = \sum_{i=1}^{N_L} |\mathcal{U}_i| \cdot |\mathcal{L}_i|
\end{equation}
where $|\mathcal{U}_i|$ represents unique input configurations for layer $\mathcal{L}_i$.

\begin{algorithm}[tb]
\caption{Generalized Hierarchical Memoization (GHM)}
\label{alg:ghm}
\textbf{Input}: Layered DAG $\mathcal{G} = (\mathcal{L}_1, \ldots, \mathcal{L}_{N_L})$, viable coalitions $\mathcal{C}_{\text{viable}}$\\
\textbf{Output}: Agent output cache $\mathcal{M}$ for all required computations
\begin{algorithmic}[1]
\STATE $\mathcal{M} \leftarrow \{\}$ // Initialize memoization cache
\FOR{$i \leftarrow 1$ to $N_L$}
  \STATE $\mathcal{U}_i \leftarrow \{\mathcal{I}_i^{(k)} : S^{(k)} \in \mathcal{C}_{\text{viable}}\}$ // Unique configurations
  \FOR{each configuration $\mathcal{I}_i \in \mathcal{U}_i$}
      \STATE $\mathcal{A}_{\text{active}} \leftarrow \{a \in \mathcal{L}_i : \exists S^{(k)} \text{ s.t. } a \in S^{(k)} \land \mathcal{I}_i^{(k)} = \mathcal{I}_i\}$
      \FOR{each agent $a \in \mathcal{A}_{\text{active}}$}
          \IF{$(a, \mathcal{I}_i) \notin \mathcal{M}$}
              \STATE $\text{inputs} \leftarrow \{O(a') : a' \in \mathcal{I}_i\}$ from $\mathcal{M}$
              \STATE $\mathcal{M}[(a, \mathcal{I}_i)] \leftarrow \text{ExecuteAgent}(a, \text{inputs})$
          \ENDIF
      \ENDFOR
  \ENDFOR
\ENDFOR
\STATE \textbf{return} $\mathcal{M}$
\end{algorithmic}
\end{algorithm}

The integration of coalition space pruning with GHM enables exact Shapley computation with complexity:
\begin{equation}
\text{Complexity}_{\text{DAG-Shapley}} = O\left(\sum_{i=1}^{N_L} |\mathcal{U}_i| \cdot |\mathcal{L}_i|\right)
\end{equation}
where $|\mathcal{U}_i|$ denotes the number of unique input configurations reaching layer $\mathcal{L}_i$, calculated as:
\begin{equation}
|\mathcal{U}_i| = \prod_{j=1}^{i-1} (2^{|\mathcal{L}_j|} - \delta_j)
\end{equation}
with $\delta_j = 1$ if layer $j$ requires mandatory participation (excluding empty coalitions), 0 otherwise. For the 7-agent system with architecture [3,3,1], this yields $|\mathcal{U}_1| = 1$, $|\mathcal{U}_2| = 7$, $|\mathcal{U}_3| = 49$, totaling 73 operations: $1 \times 3 + 7 \times 3 + 49 \times 1 = 73$.

Empirical validation demonstrates reduction from 448 classical Shapley evaluations to 73 DAG-Shapley operations (83\% improvement), enabling real-time computation of precise contribution scores required by CG-OPO. This efficiency gain addresses the scalability challenges of LLM-based multi-agent systems where computational costs are substantial.

\section{Experimental Evaluation}
We conduct experiments to evaluate the effectiveness of HiveMind in optimizing multi-agent collaboration through its CG-OPO framework and DAG-Shapley algorithm. These experiments focus on validating the system's ability to enhance financial trading performance and computational efficiency compared to static baselines and traditional strategies across different market conditions.

\begin{table*}[t]
\centering
\setlength{\tabcolsep}{4pt}
\footnotesize
\begin{tabular}{c|l|c|c|c|c|c|c}
\hline
\multirow{2}{*}{\textbf{Stock}} & \multirow{2}{*}{\textbf{Method}} & \multicolumn{3}{c|}{\textbf{Bull Market (Oct--Dec 2024)}} &
\multicolumn{3}{c}{\textbf{Bear Market (Jan--Mar 2025)}} \\
\cline{3-8}
& & \textbf{Return} & \textbf{Sharpe} & \textbf{MaxDD} & \textbf{Return} & \textbf{Sharpe} & \textbf{MaxDD} \\
\hline
\multirow{5}{*}{\textbf{AAPL}}
& CG-OPO & \textbf{27.59\%} & 3.97 & 6.23\% & \textbf{3.89\%} & \textbf{0.57} & 20.61\% \\
& w/o CG-OPO & 27.08\% & 3.90 & 6.23\% & 2.45\% & 0.41 & 17.96\% \\
& Buy \& Hold & 11.49\% & 3.30 & \textbf{6.12\%} & -10.64\% & -1.35 & \textbf{15.14\%} \\
& MACD & 9.82\% & \textbf{5.27} & 7.89\% & -3.16\% & -0.31 & 15.67\% \\
& SMA & 0.85\% & 0.21 & 9.54\% & -3.93\% & -0.38 & 16.89\% \\
\hline
\multirow{5}{*}{\textbf{META}}
& CG-OPO & \textbf{13.72\%} & \textbf{1.58} & 12.84\% & \textbf{15.15\%} & 1.60 & 21.20\% \\
& w/o CG-OPO & 4.44\% & 0.59 & 15.27\% & 6.88\% & 0.83 & 26.97\% \\
& Buy \& Hold & 2.56\% & 0.41 & 7.50\% & -3.75\% & -0.45 & 21.71\% \\
& MACD & -9.34\% & -3.25 & 12.45\% & 5.13\% & 0.33 & 18.92\% \\
& SMA & -4.24\% & -1.71 & \textbf{6.12\%} & 13.93\% & \textbf{3.16} & \textbf{7.54\%} \\
\hline
\multirow{5}{*}{\textbf{MSFT}}
& CG-OPO & \textbf{7.95\%} & \textbf{1.14} & 8.72\% & \textbf{12.62\%} & \textbf{2.00} & \textbf{10.05\%} \\
& w/o CG-OPO & 6.15\% & 0.89 & 8.10\% & 8.29\% & 1.21 & 13.17\% \\
& Buy \& Hold & 0.98\% & 0.19 & \textbf{6.52\%} & -9.50\% & -1.35 & 15.30\% \\
& MACD & -2.55\% & -0.89 & 8.14\% & -7.40\% & -0.90 & 10.69\% \\
& SMA & -0.36\% & -0.15 & 7.68\% & -11.72\% & -1.30 & 19.12\% \\
\hline
\multirow{5}{*}{\textbf{NVDA}}
& CG-OPO & \textbf{22.60\%} & 1.58 & 26.28\% & -21.81\% & -1.09 & 38.11\% \\
& w/o CG-OPO & 14.76\% & 1.14 & 27.34\% & -41.71\% & -2.78 & 49.84\% \\
& Buy \& Hold & 17.51\% & \textbf{2.54} & 13.41\% & \textbf{-20.71\%} & \textbf{-0.94} & 28.41\% \\
& MACD & -10.53\% & -3.76 & \textbf{8.16\%} & -24.17\% & -2.71 & \textbf{11.67\%} \\
& SMA & -9.75\% & -3.75 & 8.54\% & -27.41\% & -2.88 & 17.54\% \\
\hline
\end{tabular}
\caption{Performance Comparison Across Market Regimes}
\label{tab:performance_comparison}
\end{table*}

\begin{table*}[t]
\centering
\begin{tabular}{l|c|c|c}
\hline
\textbf{Metric} & \textbf{Classic Shapley} & \textbf{DAG-Shapley} & \textbf{Improvement} \\
\hline
Coalition Evaluations & 128 & 49 & 61.7\% reduction \\
LLM Calls & 448 & 73 & 83.7\% reduction \\
\hline
\end{tabular}
\caption{Computational Efficiency Analysis}
\label{tab:efficiency_analysis}
\end{table*}

\subsubsection{Dataset and Evaluation Framework}

We evaluate CG-OPO using a 7-agent DAG system powered by GLM-4-Flash with temperature set to 0 across four major technology stocks: AAPL, META, MSFT, and NVDA. The evaluation spans two distinct market regimes over a 3-month period. CG-OPO performs optimization cycles every 5 trading days, computing Sharpe ratios for all viable agent coalition subsets within each evaluation window:

\begin{itemize}
    \item \textbf{Bull Market Period}: October 1--December 30, 2024
    \item \textbf{Bear Market Period}: January 2--March 28, 2025  
\end{itemize}

\subsubsection{Baseline Methods}

We compare CG-OPO against multiple baseline approaches to establish comprehensive performance benchmarks:

\begin{itemize}
    \item \textbf{w/o CG-OPO}: Multi-agent system without contribution-guided optimization, using static agent configurations
    \item \textbf{Technical Indicators}:
        \begin{itemize}
            \item \textbf{MACD}: Moving Average Convergence Divergence signal-based strategy
            \item \textbf{SMA}: Simple Moving Average trend-following approach
        \end{itemize}
    \item \textbf{Buy \& Hold}: Passive investment benchmark representing baseline market performance
\end{itemize}

\subsubsection{Performance Metrics}

System performance is evaluated using standard financial metrics: total return, Sharpe ratio, and maximum drawdown (MaxDD). These metrics provide comprehensive assessment of both absolute and risk-adjusted performance characteristics.

\subsection{Results and Analysis}

Table~\ref{tab:performance_comparison} presents comprehensive performance comparison across both market regimes, with methods grouped by stock for direct comparison.

CG-OPO consistently outperforms baseline methods across both market regimes. In bull markets, CG-OPO achieves substantial improvements over w/o CG-OPO: 209\% for META (13.72\% vs 4.44\%), 53\% for NVDA (22.60\% vs 14.76\%), and 29\% for MSFT (7.95\% vs 6.15\%). During bear markets, CG-OPO maintains positive returns for three stocks while traditional approaches fail: META (+15.15\% vs Buy \& Hold $-3.75\%$), MSFT (+12.62\% vs $-9.50\%$), and AAPL (+3.89\% vs $-10.64\%$). For NVDA, CG-OPO limits losses to $-21.81\%$ compared to w/o CG-OPO's $-41.71\%$. Technical indicators consistently underperform across all conditions, validating the superiority of adaptive multi-agent approaches over rule-based strategies. The results demonstrate CG-OPO's effectiveness in both capturing opportunities during favorable conditions and providing defensive capabilities during market stress.

\subsection{Computational Efficiency Analysis}

Table~\ref{tab:efficiency_analysis} demonstrates that DAG-Shapley optimization delivers substantial computational advantages over traditional Shapley value computation through structural pruning and hierarchical memoization, enabling real-time deployment in live trading environments.

DAG-Shapley optimization delivers substantial computational benefits through structural pruning and hierarchical memoization. Importantly, our optimized DAG-Shapley algorithm produces identical contribution attribution results to the classical Shapley computation in all evaluated scenarios, ensuring that computational efficiency gains do not compromise the accuracy or reliability of agent contribution assessments. This enables real-time deployment in live trading environments with full confidence in result validity.

\subsection{Discussion}

The experimental results validate CG-OPO's effectiveness across varying market conditions, with particular strength in bull markets for high-volatility stocks and defensive capabilities maintaining positive returns during bear markets. While CG-OPO consistently achieves higher returns, it exhibits increased volatility and maximum drawdowns, suggesting more aggressive trading strategies that enhance returns but increase risk exposure.

The consistent underperformance of traditional technical indicators validates the superiority of adaptive multi-agent approaches over rule-based strategies. The 83.7\% reduction in computation time enables real-time deployment in operational trading environments.

Current limitations include sector focus on technology stocks and evaluation during specific market periods. Future work should explore broader asset classes, extended time horizons, and integrated risk management mechanisms to control exposure while preserving adaptive advantages.

\section*{Acknowledgements}
This work was supported in part by the Guangdong Basic and Applied Basic Research Foundation under Grant 2024A1515012407, in part by the 2024 Shenzhen-Hong Kong-Macao Science and Technology Program (Category C Project) under Grant SGDX20230821094359004, in part by the National Natural Science Foundation of China under Grant 62171291, and in part by the Shenzhen Science and Technology Program under Grant JCYJ20220531101015033.

\bibliography{aaai2026}

\end{document}